\documentstyle[11pt,aaspp4]{article}

\begin{document}

\rightline{BA-96-36}
\rightline{astro-ph/9608065}

\title{An Efficient Technique for making maps\\from Observations of the \\
Cosmic Microwave Background Radiation}
\bigskip \author{L. Piccirillo, G. Romeo, R. K. Schaefer and M. Limon}
\affil{Bartol Research Institute, University of Delaware, Newark, DE, 19716}

\begin{abstract} 
We describe a new technique for turning scans of the microwave sky  into 
intensity maps. The technique is based on a Fourier series analysis and  
is inspired by the lock-in deconvolution used in experiments which typically  
sweep the sky continuously. We test the technique on computer generated  
microwave skies and compare it to the more standard map making technique 
based on linear algebra. We find that our technique is much faster than the 
usual technique and, in addition, does not suffer from the problem of
memory limitations.  Lastly we demonstrate that the technique works under real
experimental conditions using observations of the moon.  
\end{abstract}


\section{Introduction}

One of the most important tests for models of the formation of  cosmological
structures is through the anisotropies in the cosmic microwave background
radiation (CMBR). In order to identify which model of structure formation
matches the pattern of temperature fluctuations one must compare the amplitude
of fluctuations over a range of angular scales. Early CMBR experiments
concentrated on comparing the temperatures of adjacent spots on the sky. As the
sensitivities of experimental instruments improved, it was possible to
entertain more sophisticated observing strategies. The results of the COBE
satellite (see \cite{Benn96} and references therein) and the FIRS balloon
experiment (\cite{Gang93}) have demonstrated the superiority of
making maps of the sky rather than simply finding temperature differences in a
limited number of adjacent spots. In a map, one has a large number of
temperature differences which can be found on a variety of different angular
scales, which is a much better characterization of the intrinsic pattern in the
sky than a measurement of the anisotropy on a single angular scale. The method
of map-making used by the COBE team consists of breaking up the sky into a
large number (6144) of $\sim 9$ square degree pixels and making a large set of 
linear
equations which must be inverted in order to have a map of the sky. As we would
now like to explore the anisotropy on much smaller scales, this method of
analysis would require us to break up the sky into a much larger number of
pixels and solve a much larger number of linear equations to derive a map, with
a considerable increase in computer time and memory. Recently, Wright et al.,
(1996), have suggested a new technique for producing mega-pixel maps from
differential radiometer data. These authors claim that their technique has a
computational cost that grows in the slowest possible way with increasing
angular resolution and number of map pixels. CPU time and dimension of the  RAM
is proportional to the number of pixels. We will show that our approach has a
CPU time also proportional to the number of pixels but the dimension of the RAM
is kept constant.
The inspiration for the method
to be described here comes from the fact that the lock-in deconvolution
technique - used in all sorts of differential CMBR measurements - can be used to
directly determine the Fourier components of the temperature pattern on the sky
(see also \cite{Nette96}). In this paper we develop this idea and
test it on computer generated microwave sky (with and without added noise). The
paper is organized as follows. In section 2, we describe our technique. In
section 3, we describe how we generate our microwave skies for testing. In
section 4, we compare our analysis technique with that used in making the COBE
sky maps, and show a reconstructed map of the moon made from our observing site
in Tenerife.  We end the paper with some conclusions in section 5.

\section{The Fourier approach}

We assume that our CMB anisotropy experiment is performed by throwing the beam
an angle $\theta (t)$ in the sky sinusoidally according to: 

\begin{equation}
\theta(t)= \theta_{0}\ {\rm sin}(\omega t)
\label{theta} 
\end{equation}

where $\theta_{0}$ is the maximum angular throw and $\omega/(2\pi)$ is the
oscillation frequency. The output of our detector will be the signal $S(t)$.
The fundamental idea for the inversion of data comes from the fact that given a
periodic signal $S(t)$, it is possible to express it as a Fourier expansion,
that is:

\begin{equation} 
S(\theta(t)) = {a_{0} \over 2} + \sum_{m} {a_{m} {\rm cos}(m\omega t + \phi)}+
\sum_{m} {b_{m} {\rm sin}(m\omega t + \phi)} 
\label{fourierd}
\end{equation}

where $\phi$ is a phase, usually introduced by the detector electronics, which
represents the phase delay between the sky signal and the voltage at the output
of the detector.
For sake of simplicity $\phi$ can be assumed to be
zero. The coefficients of the expansion are given by

\begin{eqnarray} 
a_{m} &=& {2\over N} \sum_{i} {s_{i} {\rm cos}(m\omega t_{i})} \nonumber \\
b_{m} &=& {2\over N} \sum_{i} {s_{i} {\rm sin}(m\omega t_{i})} 
\label{fcoef}
\end{eqnarray}

where N is the number of points sampled per cycle of the sinusoid, $s_{i}$ is
the i-th signal collected in the i-th time interval $t_{i}$. Notice that the 
above coefficients are proportional to the output components of a lock-in
amplifier using reference functions which are sines and cosines rather than a
square wave. Since we are using a set of orthonormal functions to expand our
signal, each harmonic component is independent.   The phase
$\phi$ can be chosen so that the signal S(t) can be
reconstructed by a series of sine (or cosine) terms. The orthogonal
(quadrature) components, i.e. the cosine (or sine) terms will not contain any
sky signal. It is evident from eq. 2 that the sky signal S($\theta$(t)) is the
sum of these components: 

\begin{equation} 
S(\theta(t_{i})) = {a_{0} \over 2} + \sum_{m} {a_{m} cos(m \omega t_{i}+ \phi)} 
\label{signal}
\end{equation}

\begin{equation}
R(\theta(t)) = \sum_{m} {b_{m} sin(m \omega t_{i}+\phi)}
\label{spurious}
\end{equation}

$R(\theta(t))$ contains all the spurious signals which are still coherent
with the sinusoidal beam throw but are not produced by sky sources.
Microphonics or glitches in the bolometers, for example, are easily controlled
by studying $R(\theta(t))$. 

Let us study the case of a typical ground-based
experiment where the beam is thrown sinusoidally in the sky by wobbling either
the primary or the secondary mirror of the telescope while the sky is drifting
by. Let us suppose that N points are collected during every cycle.  The sky
intensity at the positions $\theta_{i}$ along the beam throw can be obtained by
inserting in eq. \ref{signal} the times given by:

\begin{equation}
t_{i} = {1 \over \omega} {\rm arcsin} ({\theta_{i} \over \theta_{0}}).
\label{times}
\end{equation}

In summary, the basic idea consists of 1) sampling the signals from the
detectors at high speed, 2) extracting the fundamental and higher harmonics
coherent with the motion of the beam in the sky. Each of these harmonics
provides rejection to all the noise sources (mainly from the detectors) which
are not coherent, 3) making the inverse Fourier transform to recover the signal
$S(\theta)$. Notice that the absolute intensity sky is recovered along the
strip defined by the motion of the beam  modulation. Differential experiments,
with AC coupled detectors, will have $a_{0} = 0$ and a differential map is
obtained. It must be emphasized that this method is independent of the
observing pattern. Indeed, we have tried three different patterns, (see fig.
1), and all produced the same result.
        
\section{Simulating the observations}

Our aim is to show that we are capable of inverting the observational data in
order to get a map of the sky emission. We will demonstrate this by using a
synthesized microwave map of the sky. To generate the microwave skies for our
simulations, we make use of the spherical harmonic expansion of the temperature
field $T(\theta,\phi)$ on the sky

\begin{equation} 
T(\theta,\phi) = T_{0} \sum_{l,m} {a_{lm} B_{l} Y^{l}_{m}(\theta,\phi)} 
\end{equation}

where $T_{0}$ is the average background temperature, $a_{lm}$ are the
temperature anisotropy coefficients, $B_{l}$ is the expansion of the
experimental beam profile of the telescope, and $Y^{l}_{m}(\theta,\phi)$ are
the spherical harmonics, (where $\theta$ is the declination and $\phi$ is the
right ascension). The beam profile here is assumed to be Gaussian with a FWHM
of 1 degree, so $B_{l}$ is given by:

\begin{equation} 
B_{l} = {\rm exp}\left[-1/2 (l(l+1)(0.4247\ FWHM)^2)\right] 
\end{equation}

From the above relation we see that the temperature sum (eq. 7) will be cut off
for $l > 300$, so that we only include $l < 300$ in our sum. It is
computationally easier for us to generate smoothed maps for simulating the
differential measurements. We calculate all $Y^{l}_{m}(\theta,\phi)$ in steps
of 0.33 degrees in both right ascension and declination over the range $ 17 \le
\delta \le 35^{\circ}$, and $ 0 \le RA \le 360$ degrees, which would be
appropriate for a mid-latitude scan of the zenith. We will simulate our
observation on a part of this map. The $a_{lm}$ are given random phases and
Gaussian random amplitude using the variance  $\langle |a_{lm}|^2\rangle$
corresponding to the cold dark matter model, with a scale free spectrum n=1, a
Hubble constant of $H_{0} = 50$ km s$^{-1}$ Mpc$^{-1}$, and a baryon fraction 
of 5\%. The particular values of  $\langle |a_{lm}|^2\rangle$ were taken from
Schaefer and de Laix (1996). 

Once the map was obtained, we performed both a COBE like simulation,
and a simulation with our new technique. In the COBE like simulation, we have
two radiometers which perform a raster scan of the sky. While in the sky, and
the radiometers are drifting, we calculated the positions of the two horns
every 0.02 seconds, and then we read the temperature from the map. Finally, we
took the difference between the two radiometers, and we inverted these data
with the same procedure used in making the COBE maps (\cite{Line96},
\cite{Jans92}).  According to the COBE
method, we divided our patch of synthesized sky in pixels of equal area. Using
+ to denote the horn which read as positive the temperatures, and - for the
other horn, then during the n-th observation we had

\begin{equation} 
D_{n} = [T_{n+}(i) - T_{n-}(j)] 
\end{equation}

after performing N observations, then we can represent our data in matrix form,
namely

\begin{equation}
\left( \begin{array}{c} D_{1}\\ \dots\\ D_{N} \end{array} \right) = \left(
\begin{array}{cccccccc} 0  &   1   &   0   & \dots &   0   &  -1   &   0   &
\dots \\ \dots & \dots & \dots & \dots & \dots & \dots & \dots & \dots \\ 0  &  
0   &  -1   &   0   &   0   & \dots &   1   & \dots \end{array} \right) \left(
\begin{array}{c} T(1)\\ \dots\\ T(k) \end{array} \right) 
\end{equation}

where N is the number of observations, and k is the number of pixels. The above
can be rewritten more compactly as a matrix equation

\begin{equation} 
{\bf D} = {\bf M T } 
\end{equation}

Notice that the dimensions of $\bf M$ are given by the number of observations 
and the number of pixels. The previous equation can be inverted, and we find 
the vector ${\bf T}$, that is

\begin{equation} 
{\bf T} = {\bf A}^{-1} {\bf M}^{T} {\bf D} 
\end{equation}

where

\begin{equation} 
{\bf A} = {\bf M}^{T} {\bf M} 
\end{equation}

As expected, the technique was extremely accurate since the correlation between
the original map and the reconstructed map was 99.99\%. On the same map, 
we also simulated three different strategies in order to test our Fourier based
technique.  In the first one, the telescope beam moves
sinusoidally along the direction of the sky drift. In the second one, we allow
the beam to move in a direction normal to the direction of the sky drift.
Finally, we allowed the beam to describe circles in the sky. This last 
strategy has some interest for future satellite and/or balloon-borne
experiments. In each of the above cases we were able to reconstruct the
simulated map. To simulate the observations, the position (right ascension and
declination) of the center of the beam was calculated for every point, while
also allowing the sky to drift. Once the position was obtained, we read the
temperature on the map with the same right ascension and declination. In the
first simulation we used a frequency for the sinusoidal motion of 2.0 Hz, with
64 points sampled every cycle. This choice of parameters correspond to our
Tenerife 1996 bolometer campaign (\cite{picci96}).

\section{Inverting the data}

Let us now illustrate our method for the inversion of maps for each of the
three strategies of observation as shown in fig. 1. In the horizontal scan 
(fig. 1b) the
beam center moves sinusoidally along the direction of the sky drift. By using
eq. \ref{signal},  where we considered only the first 6 harmonics, we are able
to reconstruct the temperature at the beginning of every cycle, so that we get
the temperature profile along a fixed declination; then by moving the beam
center to a new declination, we obtain a two dimensional map of the sky. This
strategy has some important limitations. First of all, the bi-dimensional map
is obtained by joining strips at constant declination measured at different
times. Atmospheric and/or detector drifts can be so large that they will
destroy the  correlation between different  declinations. These problems can be
greatly reduced by using either the normal scan or the circular scan. The
reason is that these two techniques, and expecially the circular one, are
intrinsically bi-dimensional. A 2 dimensional map can be obtained in a time
scale which could be chosen to be negligible with respect to both the
atmospheric and the detector drifts. Let us first analyze the normal scan.
Suppose that our telescope beam is centered at a declination of 23 degrees,
that the peak to peak amplitude of the sinusoidal oscillation is 6 degrees and
that 64 points are sampled every cycle. By using eq. \ref{signal}, we will map
the temperature profile along different parallels from $\delta = 20^{\circ}$
to $\delta = 26^{\circ}$. 
For instance, suppose that we want to map the temperature profile at 26
degrees, then the time which enters in eq. \ref{signal} is given by $t = {16
\over 64} T$ which corresponds to the time at which the antenna is pointing at
$\delta = 26^{\circ}$. Similarly, the profile for other declinations can be
found by adjusting $t$ accordingly. The same considerations will apply to the
circular scan. 

We can now discuss the results of our map making technique. In figure 2a we
have the map used in our simulations, while in fig. 2b \& 2c we have the
reconstructed maps respectively with the COBE approach and our new technique.
It is evident that COBE method is as accurate as our new approach. From the
figures it can be seen that both techniques are extremely accurate, both
generated maps which have a correlation with the original map very close to
100\%. However, we must take into account the fact that the COBE approach has
some disadvantages: first of all, it is slow when compared with our technique:
to invert the same map it took about 4 hours of CPU time with the COBE like
inversion, while it took 20 minutes with our method. Also, the method can be
applied in real time so that it is possible to get a quick look at the maps
when the data are still in the collection phase. The second and probably most
important limitation of COBE inversion is the difficulty of obtaining high
spatial resolution maps. As said previously, COBE method involves the use of
matrices whose dimensions are given by the number of pixels and the number of
observations. An increase in the map resolution means an increase in the matrix
dimensions. The next generation of anisotropy satellite experiments expected to
achieve sub-degree spatial resolution would face  enormous computational
problems in inverting huge matrices. It is possible to consider sub-matrices so
that the matrix dimensions are reduced but the calculation involved would slow
down the computation time. Our technique, on the other  hand, does not involve
any matrices, hence is fast and the resolution that  can be realized is only
limited by the beam size.

We studied our technique in the presence of detector noise and on the higher
resolution map. We introduced the instrumental noise at every point sampled
and, for simplicity, we assumed it had a Gaussian distribution. To study a
more realistic scenario, we simulated a mid-latitude balloon-borne
experiment with bolometers having a noise of 200 $\mu K/\sqrt{Hz}$.
The instrumental beam size was 20' FWHM and the beam throw was a sinusoid
of 80' peak to peak amplitude. We simulated a 10 hours flight on a
map at 3' spatial resolution. This choice of parameters is very similar to the
MSAM-2 experiment (\cite{kowi95}). 
The original and reconstructed map are visible in fig. 3a
and 3b. Notice that the reconstructed map has also a 3' pixel size.

   Lastly, we present an elementary application of the technique to real
data.  During our recent microwave observing campaign in Tenerife,
(\cite{picci96}) we made observations of the moon for purposes of calibration.  
Due to the location of the moon in the sky over Tenerife, it was much easier to
get an accurate measurement using a horizontal scan rather than a vertical 
one.  We therefore can reconstruct a one-dimensional map of the moon using the
techniques described in this paper. (To make a two-dimensional map, we would
require additional scans at adjacent declinations.)  In figure
4, we present our reconstruction of the moon map.  As in the simulations, the
moon is smoothed by our beamsize of 1.3$^\circ$ FWHM.  Here we show the data
from one channel with $\lambda=2.1$ mm.  We use 5 harmonics in the sum in eq.
\ref{signal} to
reconstruct the moon.  For comparison we have made a simple simulation of what
signal is expected from a moon (approximated as a uniform disk) smoothed by our 
Gaussian beam.  We also express the simulation as a Fourier series with only 5
harmonics, which accounts for the apparent features at the moon's edges.  We 
see that the mapping technique works quite well in this situation under real
observing conditions.  

\section{Conclusion}

From our simulations and our simple application to real data, we have
demonstrated that this new technique, based on Fourier series works quite well. 
We learned
that the most efficient observational strategy is the one where the beam center
moves fast and explores the maximum amount of sky in the minimum time,
compatible with the speed and noise of the detectors. From the comparison with
the COBE approach, it seems that our method offers more advantages especially
in terms of computer memory limitation and speed when we want to obtain sky
maps at high spatial resolution. The relatively high speed of the algorithm
will also allow the realization of efficient quick-look programs, i.e.
real-time visualizations of the inverted data when the experiment is still in
the phase of the data collection.

\newpage

\section{FIGURE CAPTIONS}

\figcaption{The three observing strategies discussed are shown. From the top
to the bottom, they are: (a) vertical scan, (b) horizontal scan and (c) circular
scan, all centered on declination $\delta$.}

\figcaption{ The upper panel (a) is the original map. This is the computer
generated map over which we performed both our simulations of the observations.
It has already been smoothed with a Gaussian beam profile with FWHM = 1 degree.
The pixels are 1/3 of degree by 1/3 of degree and contours are labelled in 
$\mu K$. Panel (b) is the reconstructed
map by using COBE method. Panel (c) is the reconstructed map by
using our Fourier based approach, and taking into account only the first 6
harmonics. The correlations between the reconstructed 
maps and the original one are both very
close to 100 \%.}

\figcaption{ The upper panel (a) is the higher resolution original map. The
pixels are 0.05 degree by 0.05 degree. This map was used only to perform a
Fourier simulation since the number of pixels is too large to perform a
COBE-like simulation on our computer. The lower panel (b) is the Fourier
reconstructed map using 9 harmonics. This last map has been obtained by adding
an instrumental noise of 200 $\mu$K/$Hz^{1/2}$. We
simulated 10 hours of observations corresponding to a mid-latitude
balloon-flight.  The correlation between this map and the original map is about
93 \%.
}

\figcaption{ The reconstructed intensity map of the moon made using a 
horizontal scan through the center of the moon's disk with our microwave 
telescope in Tenerife using the 2.1 mm wavelength channel.  The light line is 
a simulation of the expected signal of the moon seen through a Gaussian beam 
of 1.3$^\circ$ FWHM.  Both the map and the simulated moon use only 5 harmonics 
in the Fourier series, leading to rounded moon ``edges".  }

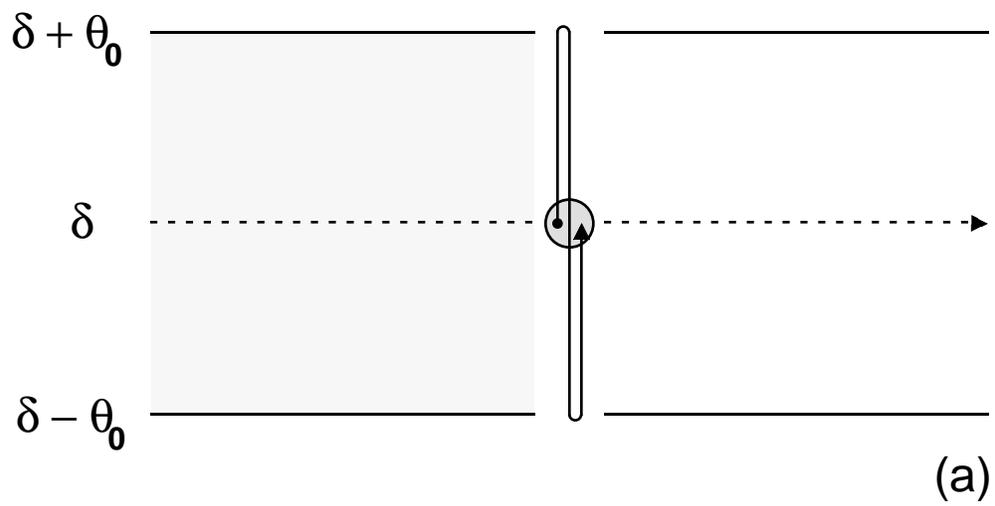
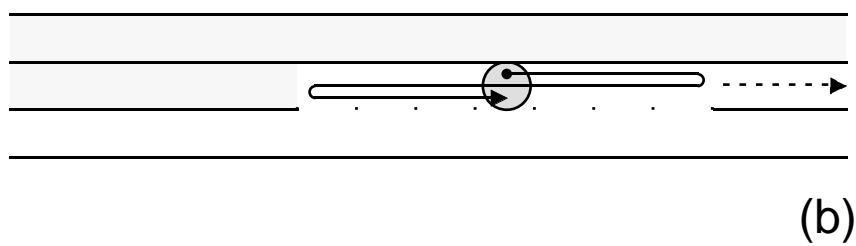
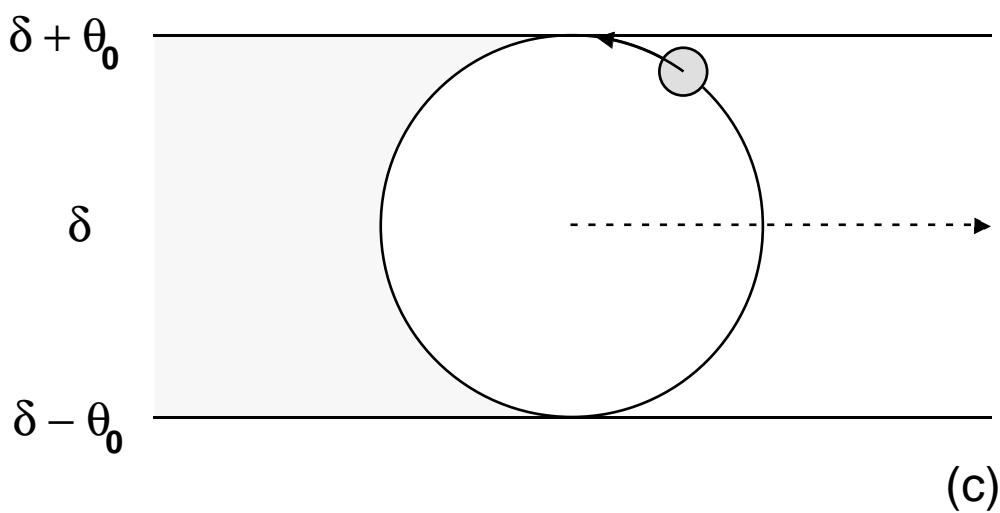
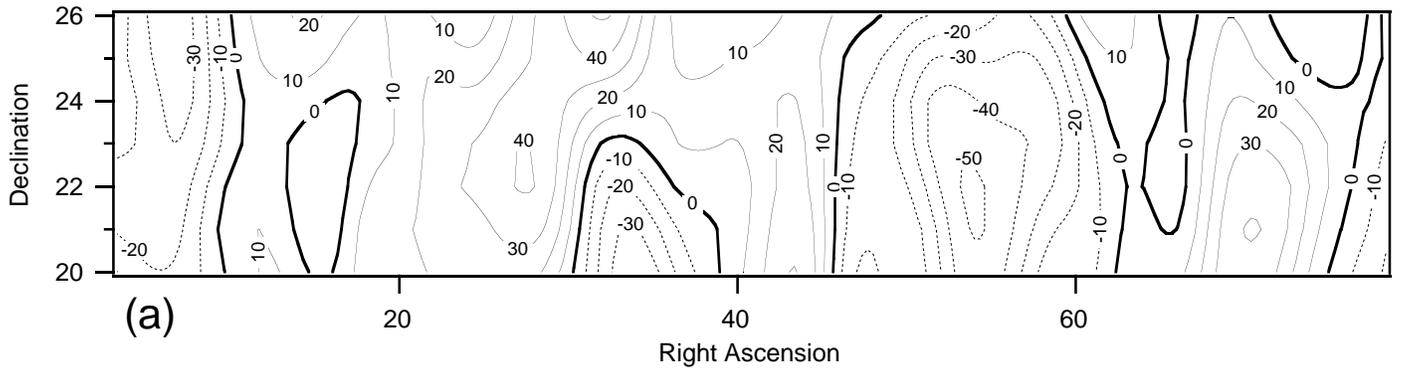
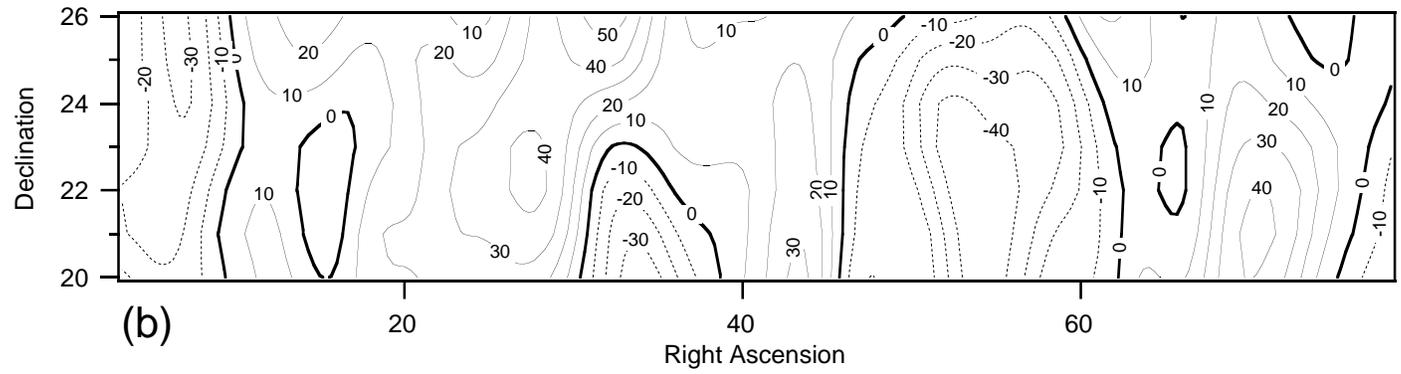
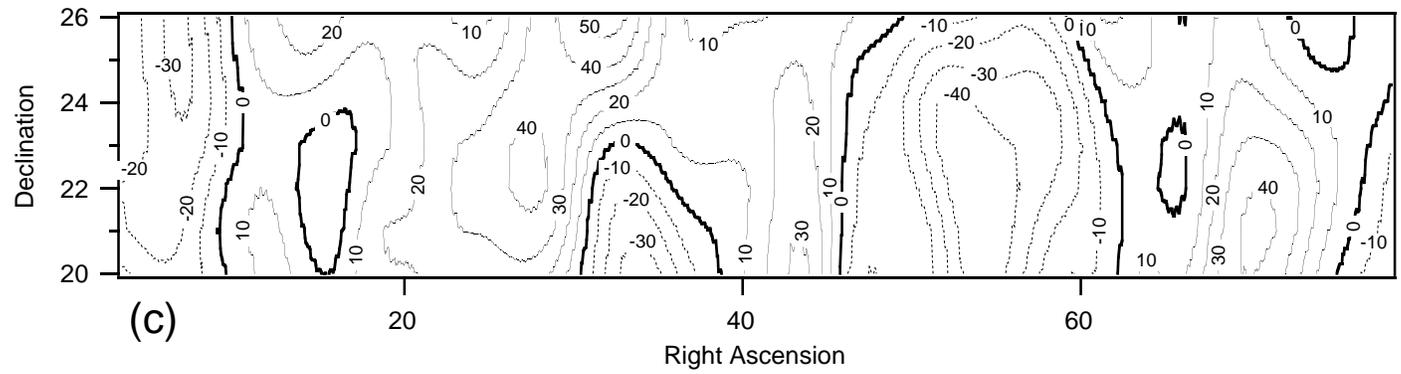
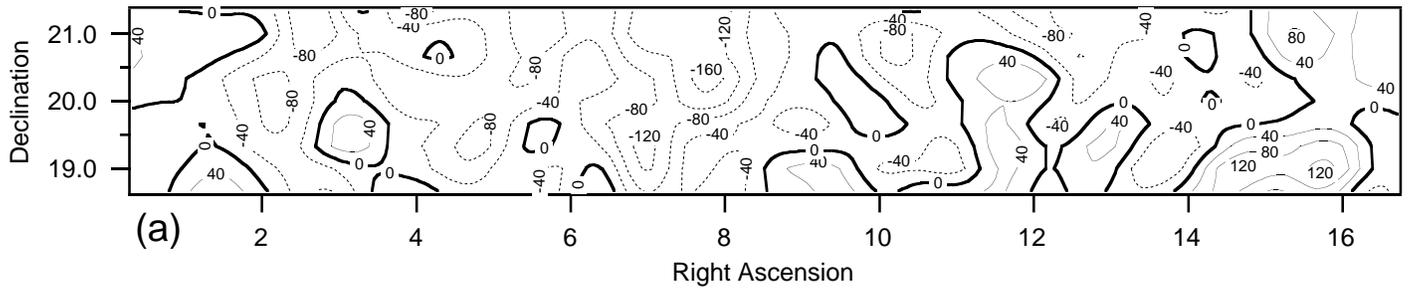
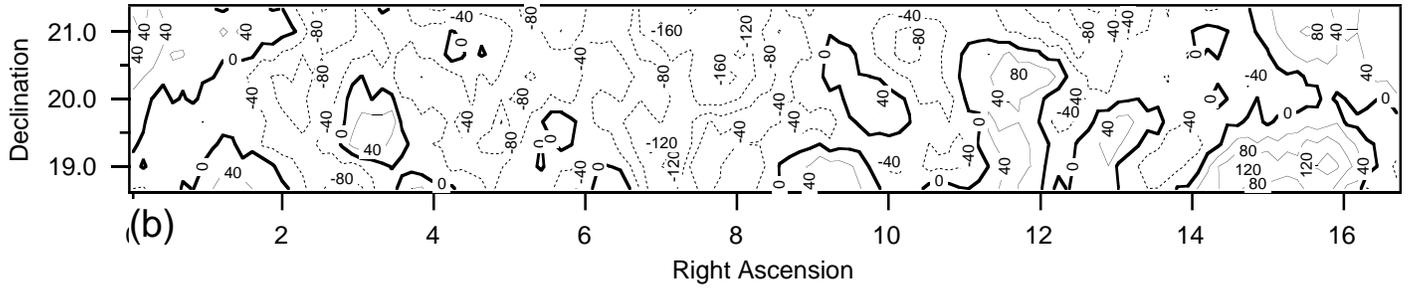
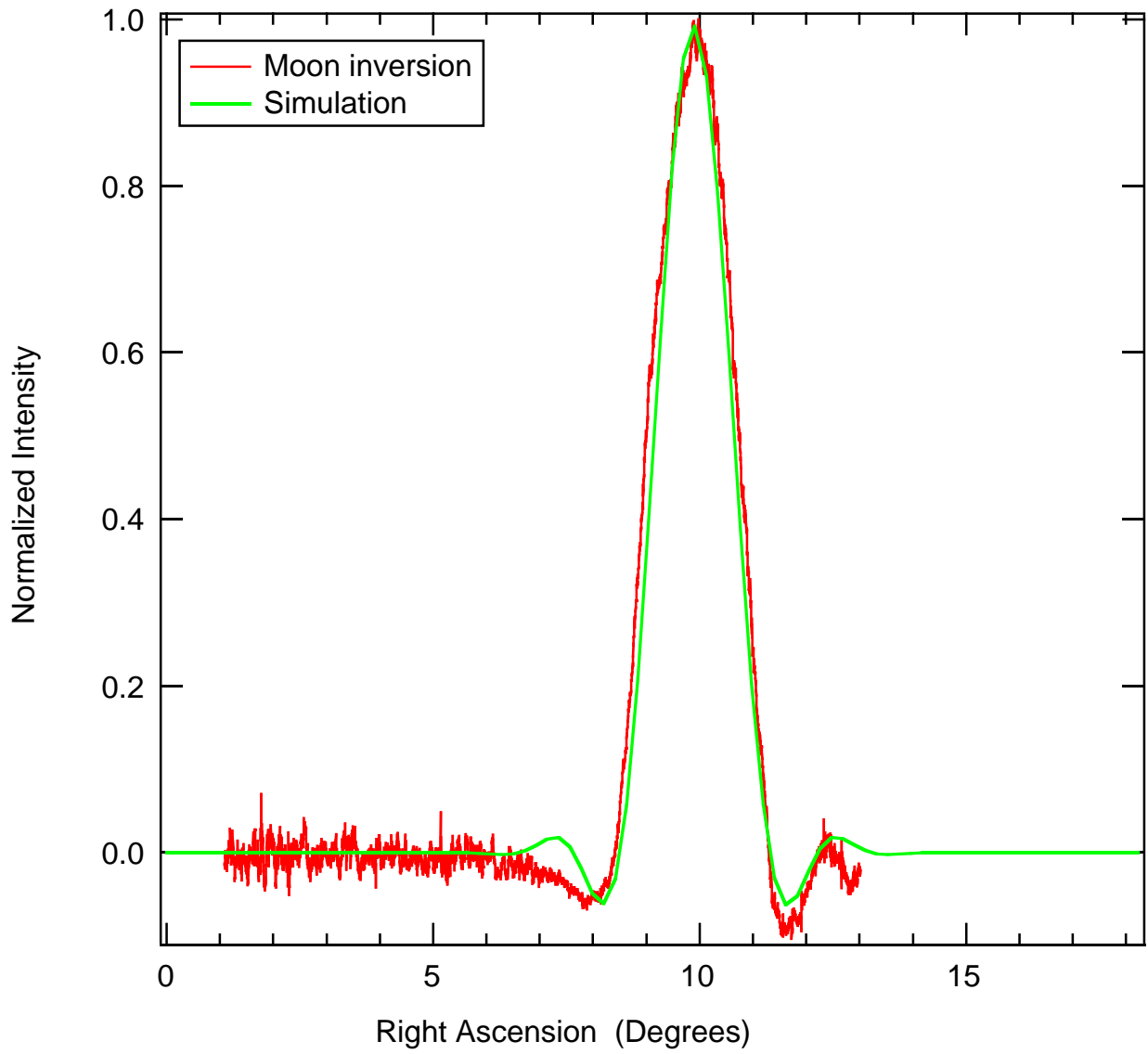
\end{document}